\documentclass[fleqn,usenatbib]{mnras}

\usepackage[T1]{fontenc}
\usepackage{ae,aecompl}

\usepackage{graphicx}	
\usepackage{amsmath}	
\usepackage{amssymb}	
\usepackage{physics}
\usepackage[swedish, english]{babel}

\newcommand{\ang}{\textup{\AA}}
\newcommand{\Msol}{M_{\odot}}

\graphicspath{ {figures/} }

\title[Star formation sampling at high redshifts]{The impact of star formation sampling effects on the spectra of lensed $z>6$ galaxies detectable with \textit{JWST}}

\author[Anton Vikaeus et al.]{
Anton Vikaeus,$^{1}$\thanks{E-mail: anton.vikaeus@physics.uu.se}
Erik Zackrisson,$^{1}$
Christian Binggeli$^{1}$
\\
$^{1}$Observational Astrophysics, Department of Physics and Astronomy, Uppsala University, Box 516, SE-751 20 Uppsala, Sweden.\\
}

\date{Accepted XXX. Received YYY; in original form ZZZ}

\pubyear{2019}

\begin{document}
\label{firstpage}
\pagerange{\pageref{firstpage}--\pageref{lastpage}}
\maketitle

\begin{abstract} 
The upcoming \textit{James Webb Space Telescope} (\textit{JWST}) will allow observations of high-redshift galaxies at fainter detection levels than ever before, and \textit{JWST} surveys targeting gravitationally lensed fields are expected to bring $z\gtrsim 6$ objects with very low star formation rate (SFR) within reach of spectroscopic studies. As galaxies at lower and lower star formation activity are brought into view, many of the standard methods used in the analysis of integrated galaxy spectra are at some point bound to break down, due to violation of the assumptions of a well-sampled stellar initial mass function (IMF) and a slowly varying SFR. We argue that galaxies with SFR$\sim 0.1\ M_\odot$ yr$^{-1}$ are likely to turn up at the spectroscopic detection limit of \textit{JWST} in lensed fields, and investigate to what extent star formation sampling may affect the spectral analysis of such objects. We use the \textsc{slug} spectral synthesis code to demonstrate that such effects are likely to have significant impacts on spectral diagnostics of, for example, the Balmer emission lines. These effects are found to stem primarily from SFRs varying rapidly on short ($\sim$ Myr) timescales due to star formation in finite units (star clusters), whereas the effects of an undersampled IMF is deemed insignificant in comparison. In contrast, the ratio between the HeII- and HI-ionizing flux is found to be sensitive to IMF-sampling as well as ICMF-sampling (sampling of the initial cluster mass function), which may affect interpretations of galaxies containing Population III stars or other sources of hard ionizing radiation.
\end{abstract}




\section{Introduction}
A standard assumption in the modelling and analysis of the spectra or spectral energy distributions of galaxies is that the stellar initial mass function (IMF) can be considered to be fully sampled, i.e., that star formation results in a sufficiently smooth distribution of stellar masses so that stochastic mass sampling effects can be ignored. Another very common assumption in this field (although rarely explicitly stated) is that the star formation rate (SFR($t$)) does not fluctuate significantly on timescales of millions of years, but can be treated as a smooth or even constant function over $\sim 10$ Myr timescales.

As has been demonstrated in a large body of work, these assumptions may be violated in low-mass systems or systems with very low star formation activity \citep[e.g.,][]{Santos97,Cervino04,Fousneau10,Fumagalli11,Eldridge12,Hernandez12,daSilva12,Paalvast17,Emami18,Krumholz19}. This is a well-known problem in the study of star clusters and dwarf galaxies in the low-redshift Universe, but has not been widely recognized as an important issue in the study of galaxies at high redshifts ($z$), since the majority of the currently observable high-$z$ objects tend to be too massive (or conversely, to have too high SFR) for star formation sampling effects to have any significant impact on their spectra. 

Even so, a few papers have highlighted the impact that star formation stochasticity may have on the interpretation of observations of Lyman-$\alpha$ equivalent widths and Lyman continuum fluxes at $z\gtrsim 3$ \citep{ForeroRomero13}, on feedback effects in simulations of the lowest-mass galaxies in the reionization epoch \citep{Applebaum18} and on the spectral signatures of low-mass Population III (Pop III) galaxies at $z\gtrsim7$ \citep{MasRibas16}. While the latter two examples deal with galaxies in a mass range that will remain well below the detection threshold of the telescopes that will come online in the foreseeable future, we here highlight that the upcoming James Webb Space Telescope (\textit{JWST}) is likely to detect large numbers of low-SFR, $z\gtrsim 6$ objects in deep observations of strongly lensed fields, for which star formation sampling effects may well have considerable observational effects. We focus on the Balmer emission lines (H$\alpha$ and H$\beta$ in particular) which provide an observed flux representative of the stars dominating the blue part of the spectrum (rest-frame UV to optical) in high-redshift galaxies. The impact of stochastic star formation on the Lyman-$\alpha$ flux from galaxies at $z\gtrsim6$ will not be studied in this article due to the considerable complexity in modeling the reprocessing of this radiation over cosmic distances.


In Sect.~\ref{motivation}, we make the case that $z\gtrsim 6$ galaxies significantly affected by star formation sampling effects are likely to turn up above the \textit{JWST} detection limit in deep observations of cluster lensing fields. Sect.~\ref{models} describes the methods and assumptions used to assess the impact of star formation sampling effects on the spectra of low-SFR galaxies at high redshift. In Sect.~\ref{SED_effects}, we explore some of the star formation sampling effects that this may have on their spectral properties. Sect.~\ref{discussion} discusses the implications of our findings.

\section{Star formation sampling effects at high redshifts with \textit{JWST}}
\label{motivation}
\begin{figure}
  \centering
    \includegraphics[width=\columnwidth]{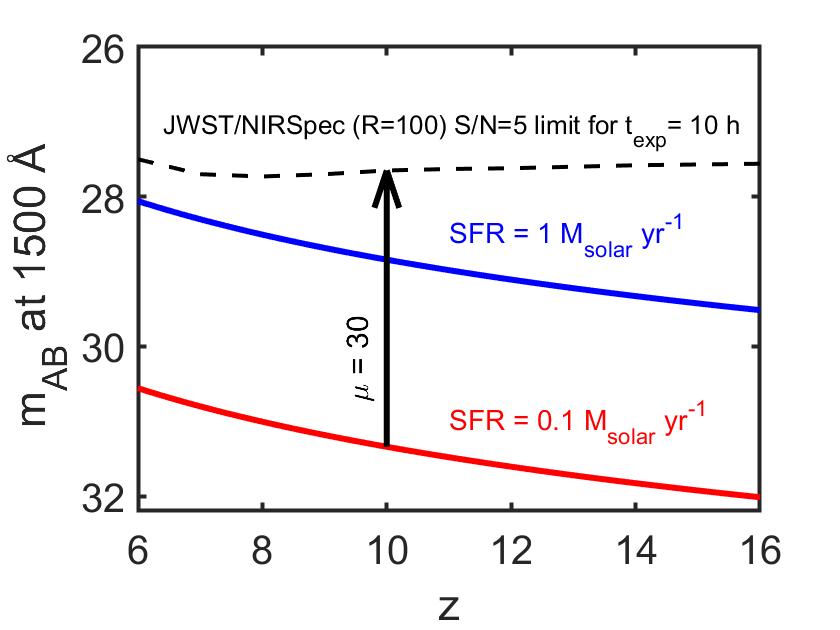}
\caption{Apparent AB magnitude at rest-frame wavelength 1500 $\ang$ as
a function of redshift for $\mathrm{SFR}=0.1\ M_\odot \mathrm{yr}^{-1}$
(red solid line) and $1\ M_\odot \mathrm{yr}^{-1}$ (blue solid) based on
the SFR-$L_\mathrm{UV}$ scaling adopted in \citet{Madau14}, rescaled to a Kroupa IMF. The dashed line
indicates the redshift-dependent \textit{JWST}/NIRSpec $R=100$, $S/N\approx 5$
point-source continuum detection limit that corresponds to a rest-frame wavelength of 1500 {\AA} for an
exposure time of 10 h. While $z\approx 6$--16 galaxies with
$\mathrm{SFR}=0.1$--1 $M_\odot \mathrm{yr}^{-1}$ are intrinsically
expected to lie below this \textit{JWST} spectroscopic detection limit, they
would be readily detectable in a strongly lensed field (as an example, the black arrow
indicates how the red line would be shifted in the case of magnification
$\mu \approx 30$).}
\label{mab_fig}
  \end{figure}

The SFR limit at which star formation sampling effects start to have significant effects on the the predictions of spectral features can be casually estimated to be $\mathrm{SFR} \lesssim 0.1\ M_\odot \mathrm{yr}^{-1}$ \citep[e.g.,][]{Paalvast17}, although some effects may be important even at $\mathrm{SFR} \sim 1\ M_\odot \mathrm{yr}^{-1}$ \citep{ForeroRomero13}. In fig.~\ref{mab_fig}, we demonstrate that the intrinsic rest-frame UV (1500 \AA) fluxes of $\mathrm{SFR}=0.1$--1 $M_\odot \mathrm{yr}^{-1}$ galaxies are expected to lie below the \textit{JWST} spectroscopic detection limit (here taken to be the \textit{JWST}/NIRSpec $R=100$, $S/N=5$, point-source continuum detection limit for an exposure time of 10 h) at $z\geq 6$. However, such galaxies are nonetheless expected to turn up in spectroscopic \textit{JWST} samples that target cluster lensing fields. In the absence of extinction, the apparent AB magnitudes of $\mathrm{SFR} = 1\ M_\odot \mathrm{yr}^{-1}$ ($0.1\ M_\odot \mathrm{yr}^{-1}$) are expected to be $m_\mathrm{AB}\approx 28$--29.5 mag (30.5--32 mag) at $z\approx 6$--16. Magnification factors of $\mu\approx 2$--6 ($\mu\approx 20$--60) would lift such objects above the spectroscopic detection limit of \textit{JWST}, and objects at $z\gtrsim 6$ with similar intrinsic magnitudes may already have been photometrically detected with the Hubble Space Telescope in cluster lensing fields \citep[e.g][]{Yue18,Kikuchihara19}. Dust attenuation could potentially push the $\mathrm{SFR}\approx 0.1$--$1\ M_\odot \mathrm{yr}^{-1}$ objects further below the detection threshold, but simulations predict that the UV attenuation of such faint $z\gtrsim 6$ galaxies in many cases will be $<0.5$ mag \citep{Shimizu14}.

\section{Models}
\label{models}
As explained by \citet{daSilva11}, there are two entangled star formation sampling effects that become relevant at low SFRs -- the sampling of the stellar initial mass function (IMF) and the temporal variations of the SFR which are induced by sampling of the initial cluster mass function (ICMF).

The first effect stems from the fact that a coeval population of stars forming with an integrated mass below a certain limit will experience an incompletely sampled IMF \citep{Elmegreen2000}. The result of this is that stellar populations with the same age and integrated mass nonetheless may exhibit different observable characteristics due to stochasticity in the number of massive stars present. Since the mass of the galaxy at any time is a result of its star formation history, this also translates into a corresponding limit on SFR($t$) -- an object with lower SFR($t$) will produce a lower integrated mass over a given time span and hence be more prone to an undersampled IMF. In this article, the effects regarding an undersampled IMF will be referred to as IMF-sampling.

The second effect stems from the assertion that if stars form in coeval star clusters sampled from an initial cluster mass function, the SFR($t$) will inevitably fluctuate on short timescales, as any galaxy that happens to form a very massive cluster will see a momentary peak in its temporal SFR. As a result, galaxies with the same mean SFR (when measured over timescales longer than a few Myr) will exhibit different observed characteristics depending on whether they happen to be experiencing a temporary peak or a drop in the stochastically fluctuating SFR($t$). In this article we will refer to the stochastic behaviour due to sampling from the ICMF, simply as ICMF-sampling. When studying high-redshift galaxies and their behaviour over time, it is important to make a distinction between low-SFR and low-mass galaxies, which do not necessarily correlate, since a high-mass galaxy can have a low SFR($t$) and vice versa. Therefore, for any galaxy mass, we may very well find sampling issues. However, the relative impact of such sampling issues becomes stronger in low-mass systems, due to feedback effects during star formation -- making the low-mass galaxies more relevant when studying sampling issues. Furthermore, we must note that at high redshifts it is necessary that galaxies have ongoing star formation in order to produce any observable H$\alpha$ and/or H$\beta$ emission lines at all. Without young and massive stars that provide the hard ionizing flux needed to excite the relevant Balmer emission lines, we cannot study high-redshift galaxies in the amount of detail desired. Therefore, in this analysis, SFR($t$) is more relevant than the total mass of the galaxy since it tracks the stars responsible for the outgoing ionizing flux over time in more detail.

Here, we have used the stellar population synthesis software \textsc{slug} (stochastically lighting up galaxies; \citealt{daSilva12,Krumholz15})  
to simulate the stochasticity due to star formation sampling effects on the equivalent width (EW) of the Balmer emission lines H${\alpha}$ and H${\beta}$, the AB-magnitude $m_{\text{UV}}$ (which we define by the flux density at rest-frame wavelength 1500 $\ang$) and the $L_{\text{H} {\alpha}}/L_{\text{UV}}$ (dimensionless) ratio. The Monte Carlo based simulations are run with 100 trials each for a mean SFR of either 1 or 0.1 $M_\odot$ yr$^{-1}$ (corresponding to the two coloured lines shown in fig.~\ref{mab_fig}) for a galaxy with an age of 100 Myr and a fixed metallicity of $Z=0.004$. We adopt the \citet{kroupa2001} universal IMF in the 0.1--100 $M_\odot$ mass range, and a power-law ICMF with an exponent $\beta=-2$ throughout the 20--10$^7$ $M_\odot$ cluster mass range \citep{LadaLada2003,daSilva11,Fall2010}. We furthermore assume that all star formation takes place in clusters. We use the thermally pulsating Padova AGB stellar tracks for all calculations and base the predictions for the nebular continuum and emission line strengths on the analytical approximations implemented in \textsc{slug}, assuming no loss of ionizing photons due to dust or leakage into the intergalactic medium. To convert intrinsic luminosities to apparent AB magnitudes we adopt the $\Lambda$CDM cosmology with $\mathrm{H_0=67.6\ km \, s^{-1}Mpc^{-1}}$, $\Omega_{\Lambda}=0.69$ and $\mathrm{\Omega_{m}=0.31}$ \citep{Planck16}.

\section{Star formation sampling effects on spectral properties} \label{SED_effects}

\subsection{Effects of stochasticity}
\begin{figure*}
  \centering
    \includegraphics[width=\columnwidth]{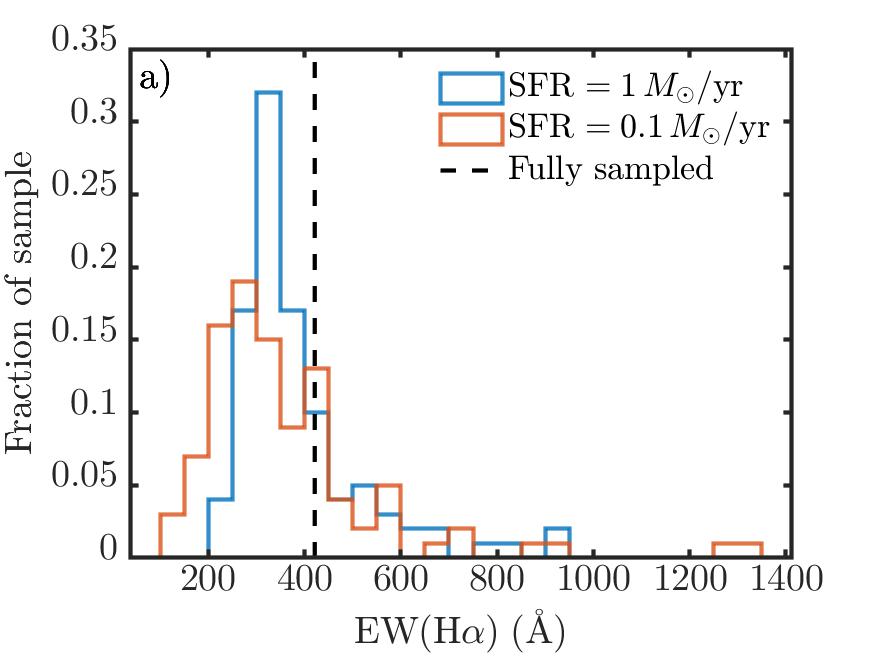}
    \includegraphics[width=\columnwidth]{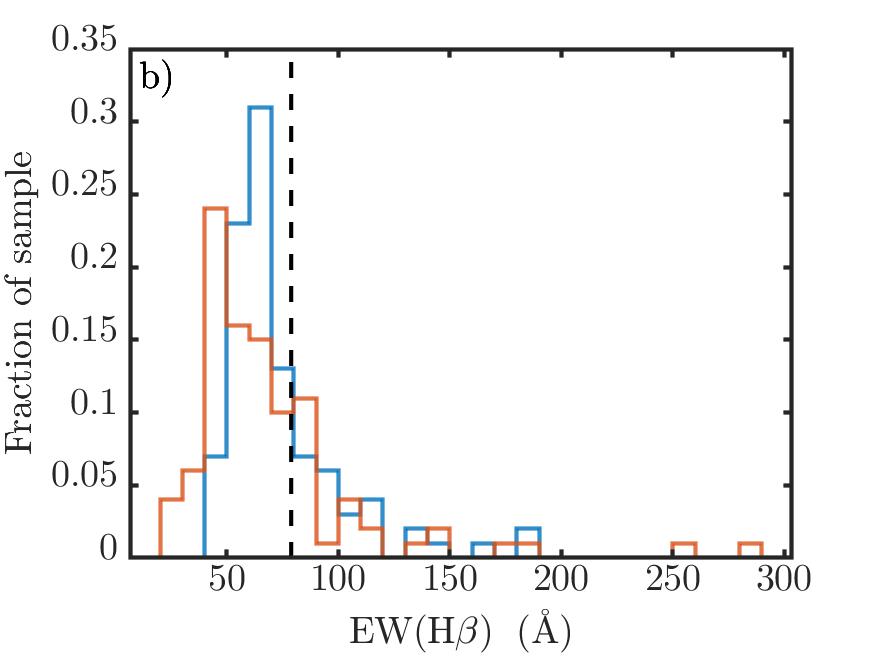}\\
    \includegraphics[width=\columnwidth]{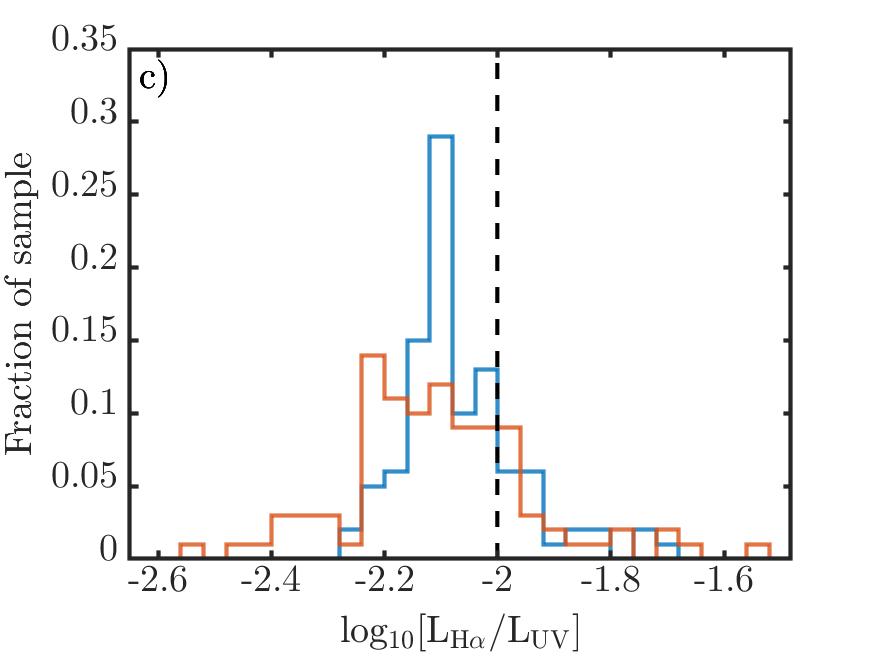}
    \includegraphics[width=\columnwidth]{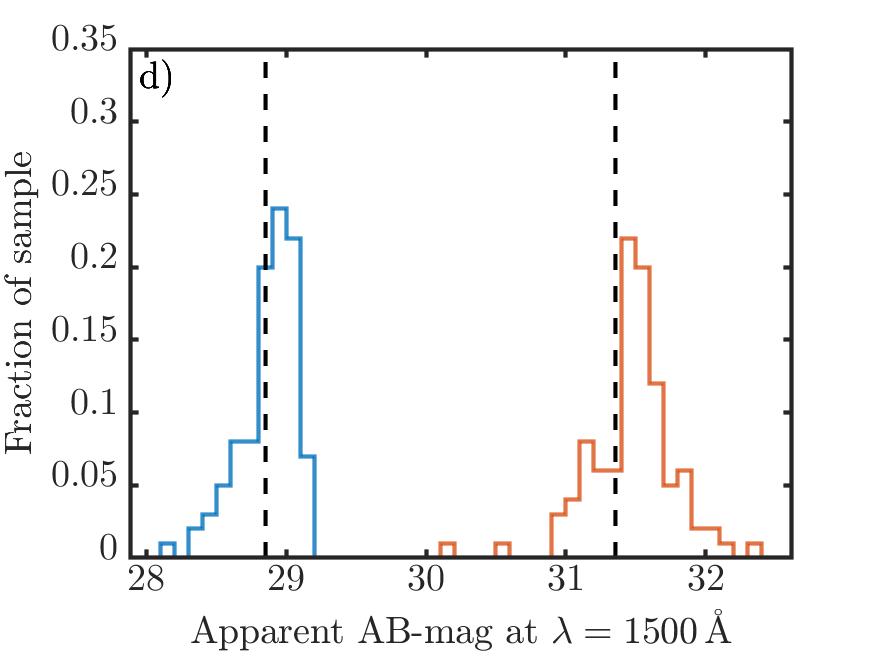}
    \caption{Distribution of rest-frame EW(H$\alpha$), EW(H$\beta$), $L_{\text{H}\alpha} / L_{\text{UV}}$ (dimensionless) and the apparent AB magnitude at redshift $z=10$ shown for 100 simulations of a 100 Myr old galaxy at SFR $= 1.0 \, \Msol$ yr$^{-1}$ (blue) and SFR $= 0.1 \, \Msol$ yr$^{-1}$ (orange) . The dashed black line represents the outcome of a fully sampled (non-stochastic) simulation. The UV is defined to be at rest-frame $\lambda = 1500 \ \ang$.}
     \label{fig2}
  \end{figure*}

In fig.~\ref{fig2}, we show the predicted effects of star formation sampling stochasticity on a number of observables relevant for high-redshift objects with \textit{JWST}: the rest-frame equivalent widths  EW(H$\alpha$) and EW(H$\beta$), the $L_{\mathrm{H}\alpha}/L_\mathrm{UV}$ ratio and the rest-frame UV apparent AB magnitude $m_\mathrm{AB}$. Note that the emission lines H$\alpha$ and H$\beta$ will be observable to redshifts of $z\approx 7.1$ and $z\approx 9.9$ respectively with \textit{JWST}/NIRSpec, while the UV (defined at 1500 \AA) will formally remain in the \textit{JWST}/NIRSpec wavelength window up to $z \approx 34$ (i.e., in an epoch prior to the formation of the first galaxies). All distributions of the observables in fig.~\ref{fig2} show a clear scatter, which is especially prominent in the case of the SFR $\approx 0.1 \, \Msol$ yr$^{-1}$ galaxies. We note that significant stochastic behaviour is also seen at SFR $\approx 1 \, \Msol$ yr$^{-1}$, but the number of extreme outliers is lower. 

The EWs of Balmer emission lines can be used as a combined constraint on age and star formation history \citep[e.g.,][]{Zackrisson05,ostlin08}, but also in principle to single out objects with extreme levels of leakage of ionizing photons into the intergalactic medium \citep[giving very low EW:][]{zackrisson13}, to find objects with extreme ionizing fluxes, like Population III-dominated galaxies \citep[giving very high EW:][]{Inoue11} or to provide a general test of the viability of stellar population models for high-redshift galaxies \citep{Zackrisson2017}. However, a significant stochastic variation in the EWs of Balmer emission lines for reasons that are largely disconnected from age or average star formation properties would significantly obfuscate such attempts. Looking at the equivalent widths of the H${\alpha}$ and H$\beta$ emission lines (fig.~\ref{fig2}a \& b), we find that the EW of the outliers is up to a factor of $\sim 4$ higher than the mean of the distribution at $\mathrm{SFR} \approx 0.1 \ M_\odot$ yr$^{-1}$ and a factor of $\sim 2$ higher at $\mathrm{SFR} \approx 1 \ M_\odot$ yr$^{-1}$. Moreover, the modes of the distributions predicted in the case of stochastic star formation sampling are centered at slightly lower EWs with respect to the value obtained when using the corresponding non-stochastic treatments. We find that the EWs of H$\alpha$ and H$\beta$ correlate almost perfectly, so that objects that are EW outliers in one line will be outliers in the other as well. The non-stochastic calculations (represented by the dashed black lines in fig.~\ref{fig2}) are calculated with \textsc{slug} by turning off cluster formation, while populating the whole galaxy according to the analytical expression for the Kroupa IMF.

As the luminosity of the H$\alpha$ emission line and the UV continuum trace star formation activity on somewhat different timescales \citep[e.g.,][]{Boquien14}, the $L_{\mathrm{H}\alpha}/L_\mathrm{UV}$ ratio has a long tradition of being used as a probe of the burstiness of stellar populations \citep[e.g.,][]{Madau14,Shivaei15}, although it can also be used to probe dust attenuation, the stellar initial mass function and the escape of ionizing photons \citep[e.g.,][]{Emami18}. Fig.~\ref{fig2}c shows the distribution of the $L_{\text{H} {\alpha}}/L_{\text{UV}}$ ratio, where $L_{\text{UV}}$ is defined as $\lambda L_{\lambda}$ (so that $[L_{\mathrm{UV}}]=$ erg/s) at a rest-frame $\lambda = 1500 \ \ang$. The overall behaviour from the Balmer EW histograms is largely reproduced, in the sense that the SFR $\approx 0.1 \, \Msol$ yr$^{-1}$ scenario results in more outliers than the SFR $\approx 1 \, \Msol$ yr$^{-1}$ one, and that the mode of the distribution lies lower than the expectation from a model that ignores the sampling effects. Moreover, we find that there is a correlation between $L_{\mathrm{H}\alpha}/L_\mathrm{UV}$ and the Balmer line EWs, in the sense that objects that are EW outliers in most cases also are $L_{\mathrm{H}\alpha}/L_\mathrm{UV}$ outliers as well (with strong EW giving high $L_{\mathrm{H}\alpha}/L_\mathrm{UV}$ ratios, and vice versa).

The UV continuum flux is often used as an SFR indicator \citep[e.g.,][]{Madau14} and also serves as an important discriminator on the overall detectability of high-redshift galaxies using either photometry or low-resolution spectroscopy. In fig.~\ref{fig2}d, we show UV magnitude ($\mathrm{m_{AB}}$  at $1500 \ \ang$) distributions obtained for 100 Myr old galaxies with $\mathrm{SFR \sim 0.1}\ M_\odot$ yr$^{-1}$ and $\mathrm{SFR \sim 1}\ M_\odot$ yr$^{-1}$ at $z=10$. The modes of the distributions agree reasonably well with the fluxes predicted in the absence of stochastic star formation sampling effects, but rare outliers that deviate up to a magnitude or more are seen at $\mathrm{SFR} \approx 0.1\ M_\odot$ yr$^{-1}$. A correlation with other spectral properties is present, as the faintest objects also tend to have very low Balmer line EWs, and the brightest ones very high. The \textit{JWST}/NIRCam instrument is expected to be able to photometrically detect objects down to $m_\mathrm{AB}\approx 28.8$ mag in 10 h (point-source limit, $S/N\approx 5$), which is very close to the expected magnitudes of $\mathrm{SFR} \approx 1\ M_\odot$ yr$^{-1}$ objects at $z\approx 10$. Similarly deep observations of a lensed field with $\mu\gtrsim 10$, will probe objects with $\mathrm{SFR} \lesssim 0.1\ M_\odot$ yr$^{-1}$ where stochastic effects start to become more severe. As low-SFR objects are more common than high-SFR ones, one may also expect to observe galaxies with even lower mean SFR which -- due to sampling effects -- attain magnitudes far brighter than one would normally predict for low-SFR objects. 

  \begin{figure}
  \centering
    \includegraphics[width=\columnwidth]{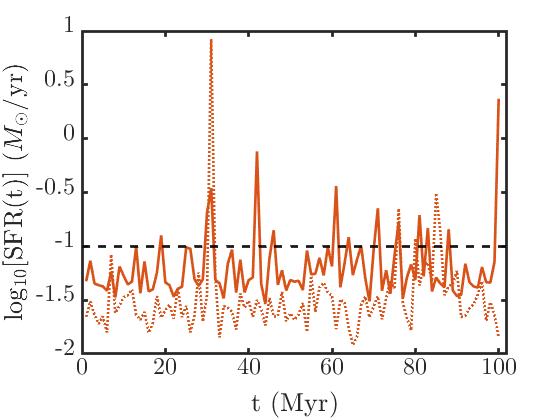}
    \caption{Actual SFR($t$) for a target SFR = $0.1 \, \Msol$ yr$^{-1}$ in the two extreme cases of the simulated galaxies. The galaxy with the strongest Balmer line equivalent widths correspond to the solid orange line while the dotted orange line represents the galaxy with the weakest Balmer emission lines. The dashed black line shows the target SFR. The extreme properties of the galaxy with the strong equivalent width are due to a large spike in the SFR at an age close to 100 Myr, which is caused by the formation of an unusually massive star cluster. Correspondingly, the galaxy with the weakest Balmer line equivalent widths is a result of a deficit of young and massive star clusters being formed at the time of emission of radiation.}
     \label{fig4}
  \end{figure}
\subsection{Sources of stochasticity}
  \begin{figure*}
  \centering
    \includegraphics[width=\columnwidth]{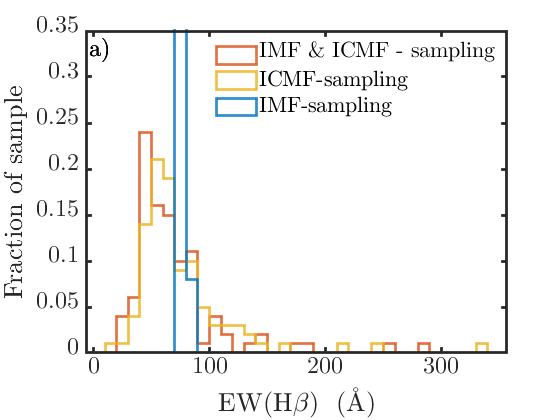}
    \includegraphics[width=\columnwidth]{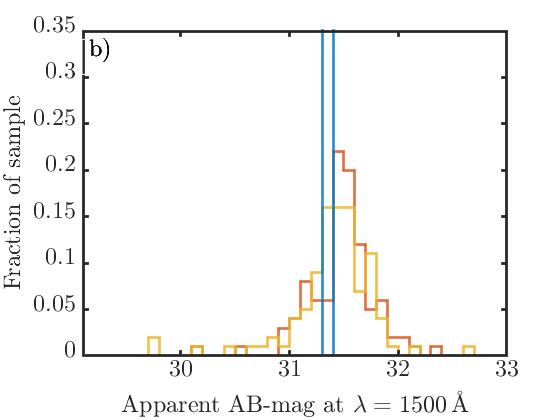}
    \caption{Distribution of a) rest-frame EW(H$\beta$) and b) apparent AB-magnitude at redshift $z=10$ for 100 simulations of a 100 Myr old galaxy with SFR = $0.1\ M_\odot$ yr$^{-1}$. The three distributions in each graph show simulations when only IMF-sampling or only ICMF-sampling is at work, as well as the situation when both IMF and ICMF-sampling are combined. The combined IMF and ICMF-sampling displayed here is the same as in fig.~\ref{fig2} (b,d) for SFR = $0.1\ M_\odot$ yr$^{-1}$.}
     \label{fig3}
  \end{figure*}
  
  \begin{figure*}
  \centering
    \includegraphics[width=\columnwidth]{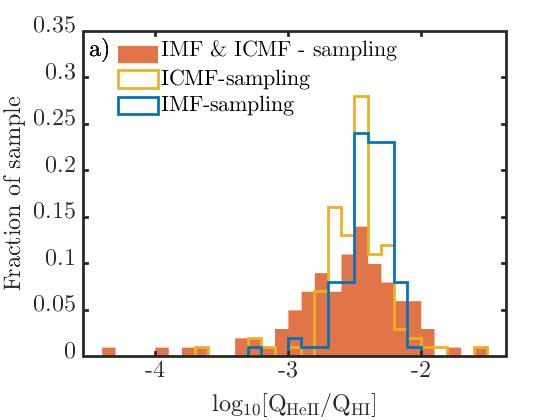}
        \includegraphics[width=\columnwidth]{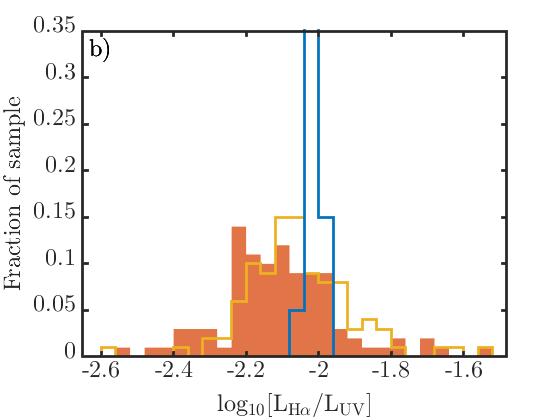}
    \caption{Distribution of a) rest-frame $Q_\mathrm{HeII}/Q_\mathrm{HI}$ and b) $L_{\text{H}\alpha} / L_{\text{UV}}$ (dimensionless) at redshift $z=10$ for 100 simulations of a 100 Myr old galaxy with SFR = $0.1\ M_\odot$ yr$^{-1}$. The the distributions in each graph show simulations when only IMF-sampling or only ICMF-sampling is at work, as well as the situation when both IMF and ICMF-sampling are combined. The combined IMF and ICMF-sampling for $L_{\text{H}\alpha} / L_{\text{UV}}$ displayed here is the same as in fig.~\ref{fig2}c for SFR = $0.1\ M_\odot$ yr$^{-1}$.}
     \label{fig5}
  \end{figure*}
  
Are these stochastic effects due to IMF-sampling or ICMF-sampling? In fig.~\ref{fig3}, we attempt to disentangle the two effects in order to show that IMF-sampling alone gives rise to very modest effects on EW(H$\beta$) and the UV continuum luminosity and that the effects we have discussed are almost entirely due to ICMF-sampling. Naturally, in order to only simulate ICMF-sampling, we turned off IMF-sampling (and vice versa to simulate IMF-sampling only). In doing so one needs to acknowledge that turning off either sampling source will have effects on the other. For example, when only simulating IMF-sampling, the star formation rate is kept at a constant level of 0.1 or 1.0 $\ M_\odot$ yr$^{-1}$ (the two SFRs studied in this paper). On the other hand, when ICMF-sampling is activated, the temporary SFR$(t)$ can become significantly lower (higher) which would result in stronger (weaker) IMF-sampling effects since it is more severe on smaller populations of stars. This illustrates the fact that the sources of the variations seen in e.g., fig.~\ref{fig2} cannot be disentangled perfectly using our method since we are not running the IMF-sampling simulation on exactly the same sizes of star clusters as when ICMF-sampling is also active. What we manage to do here is rather to get a handle on which effect that seems to be the dominant one.

Examination of the star formation history of the most extreme outliers in the histograms of fig.~\ref{fig2} confirm that these are due to recent SFR spikes or troughs. As an example, in fig.~\ref{fig4} we show the star formation history for the $\mathrm{SFR} \approx 0.1\ M_\odot$ yr$^{-1}$ galaxy that has given rise to the highest Balmer line EWs, $L_{\mathrm{H}\alpha}/L_\mathrm{UV}$ ratio and brightest UV flux (orange solid line). This galaxy has experienced a recent burst of star formation, due to the formation of an unusually massive star cluster, by a factor of $\approx 20$ above the baseline SFR (the SFR around which the lines fluctuate up and down) -- temporarily reaching SFR $\approx 2\ M_\odot$ yr$^{-1}$. Galaxies that show very low Balmer line EWs also feature temporary SFR peaks, but happens to be in a low-SFR state at the time when the spectrum is extracted (orange dotted line). The two lines in fig.~\ref{fig4} have slightly different baseline SFR while they both have the same SFR when averaged over 100 Myr. The reason that the baseline SFR is lower for the galaxy belonging to the dotted orange line is simply because a very massive cluster was formed at $\sim$ 30 Myr (temporarily reaching SFR $\approx 10\ M_\odot$ yr$^{-1}$). This forces the baseline SFR down in order to achieve an $\mathrm{SFR} \approx 0.1\ M_\odot$ yr$^{-1}$, when averaged over 100 Myr.

Importantly, we should note the fact that the most extreme outliers form at a temporary peak in the SFR, when the galaxy is at its brightest. This implies that even though the outliers are more rare objects, this effect introduces a bias due to the use of flux-limited samples, that may cause them to be observed more frequently than expected. Consequently, this also implies that these objects will have formed a very massive cluster adjacent in time to when the light that we observe was emitted -- meaning that these objects will suffer less from IMF-sampling effects. On the other hand, this mentioned bias will be counteracted, on statistical grounds, due to gravitational lensing which tends to favour the most common objects. A thorough calculation of the probabilities of observing a high-redshift galaxy depends on the apparent brightness (which we have seen can vary stochastically) and number density of galaxies as well as on magnification probabilities. Given that most galaxies of a given mean SFR are not outliers, but rather have magnitudes close to the mean of the distribution, there is a larger probability for a galaxy with average spectral features being magnified, and therefore have an increased chance of being observed. What this means is that even though stochastic effects increases the probability of observing outliers in the distributions acquired in this paper, magnification effects may boost the brightness of galaxies with average spectral features and therefore we are not guaranteed to observe unexpectedly large numbers of outlier objects in future surveys. Exact determination of the probability of detecting high-redshift galaxies as a function of the galaxies SFR is highly complex and is therefore left for further studies.

Having concluded that IMF-sampling has very modest effects on EW(H$\beta$), $m_{\text{UV}}$ and $L_{\text{H}\alpha} / L_{\text{UV}}$ (displayed in fig.~\ref{fig3} and fig.~\ref{fig5}b) we turn to the ratio between the HeII- and HI-ionizing flux (defined as $Q_\mathrm{HeII}/Q_\mathrm{HI}$, where $Q$ is in units of photons/s) which is more sensitive to IMF-sampling due to the fact that the origin of strong HeII lines are the most massive stars. A galaxy populated with IMF-sampling activated will introduce scatter in the $Q_\mathrm{HeII}/Q_\mathrm{HI}$ distribution as IMF-sampling may produce clusters with widely differing numbers of stars at the highest masses. Since this ratio may be used in the determination of Pop III observational signatures through the appearance of strong HeII lines \citep[that can be used to discriminate between chemically pristine and metal-enriched stellar populations:][]{raiter10}, it is important to know whether the strong HeII lines originate from a chemically enriched stellar population subject to IMF-sampling or from a galaxy containing a combination of Pop I/II and Pop III stars \citep[the type of object predicted in the simulations of][]{Sarmento2018,Sarmento2019}. The ratio $Q_\mathrm{HeII}/Q_\mathrm{HI}$ (fig.~\ref{fig5}a) shows a notable scatter in the distribution when including IMF-sampling in the simulation which, as mentioned, differs from what we see in the quantities displayed in fig.~\ref{fig2}, where the source of the scatter is entirely dominated by ICMF-sampling. Note that we do not display EW(H$\alpha$) when separating IMF and ICMF-sampling effects, since EW(H$\alpha$) shows the same modest effects from IMF-sampling as EW(H$\beta$).

A measure of the amount of scatter in the distributions is the relative standard deviation $\sigma / \mu$ (the standard deviation divided by the mean of the distribution). In the case of IMF-sampling, the distribution of $Q_\mathrm{HeII}/Q_\mathrm{HI}$ gives a value $\sigma / \mu \sim 0.36$ while all other distributions show a value for $\sigma / \mu$ that is (at the least) an order of magnitude smaller, confirming that $Q_\mathrm{HeII}/Q_\mathrm{HI}$ is much more sensitive to IMF-sampling. Looking at the distribution of $Q_\mathrm{HeII}/Q_\mathrm{HI}$ seen in fig.~\ref{fig5}a one would, in most cases, require higher values for this ratio in order to interpret this as clear-cut Pop III signatures (no metal lines and very large values for $Q_\mathrm{HeII}/Q_\mathrm{HI}$). If the simulated sampling effects would have been larger, producing even higher values for $Q_\mathrm{HeII}/Q_\mathrm{HI}$, one would encounter a considerable risk of erroneously interpreting the flux from chemically enriched stars (with low metallicity) suffering from such sampling issues, as coming from Pop III stars. As mentioned, in most cases the $Q_\mathrm{HeII}/Q_\mathrm{HI}$ ratio found in this paper is not large enough to cause serious concerns with future attempts of detecting metal-free galaxies (i.e., galaxies containing only Pop III stars) -- only in the few cases where $\mathrm{log_{10}}[Q_\mathrm{HeII}/Q_\mathrm{HI}]>-2$ do we reach a regime consistent with some suggested Pop III galaxy models \citep[e.g.,][]{Schaerer2003}. \cite{Inoue11} showed that it is practically very hard to distinguish a galaxy containing very small amounts of metals from a zero metallicity galaxy. In a scenario where observations cannot discern any metals, and the galaxy simultaneously show very high values for $Q_\mathrm{HeII}/Q_\mathrm{HI}$, we will not be able to say with certainty whether we are observing a low-metallicity galaxy with some fraction of Pop III stars, or just a Pop II galaxy with very low metallicity, suffering from the sampling effects presented in this paper.

In fig.~\ref{fig5}a, we see that the outliers with very low values are more common, nevertheless, we find that the distribution show some outliers that reach values a factor of $\sim 6$ larger than the mean. This kind of uncertainty would imply a significant scatter in the corresponding EW of the HeII emission lines (e.g., HeII $\lambda$1640 \AA \ and $\lambda$4686 \AA). Similar discrepancies (with high HeII emission line EW) has been seen between models and observations in e.g., \citet{shirazi2012,Berg2018, Schaerer2019}, which account for several possible sources that could be responsible for the production of strong HeII emission lines, such as; Wolf-Rayet stars, active galactic nuclei, X-ray binaries.

In \cite{Sarmento2018,Sarmento2019}, it is suggested that a significant amount of Pop III flux-dominated galaxies will appear with $\mathrm{m_{UV}} \approx 31.4$ at redshift $z=9$. Based upon the scatter found in $Q_\mathrm{HeII}/Q_\mathrm{HI}$ one can argue that when using the suggested HeII 1640 $\ang$ emission line to constrain the Pop III content of such galaxies, IMF-sampling could be an important source of uncertainty that needs to be considered in order not to misinterpret the results in observations of low metallicity galaxies at the detection threshold of \textit{JWST}.

\section{Discussion}
\label{discussion}
The main conclusion of this work is that fitting the spectral properties of lensed high-redshift dwarf galaxies at the \textit{JWST} detection threshold to models that assume a smoothly varying SFR over more than a few Myr may give rise to biased results. The size of such biases can be estimated by fitting such models to mock spectra from galaxy formation simulations that simultaneously take star formation sampling effects, bursty star formation histories and metallicity distributions into account, which we leave for future work. In this paper we have simulated very young galaxies which have only formed stars continuously for $\sim$ 100 Myr, which allowed us to study spectral features at the high redshift frontier $z \sim 10$. Older galaxies which have formed stars for $\gtrsim$ 500 Myr (representative of $z\sim 6$) will also suffer from the sampling issues discussed in this paper, but to a smaller extent. Simulating the galaxies to later times resulted in lower Balmer line EWs with less scatter in the distributions, while the age of the galaxy had very modest effects on the distribution of the UV luminosity at 1500 \AA. The decreasing EW can be explained by considering the evolution of the luminosity of the emission lines relative to the continuum at the corresponding wavelengths. The average luminosity of the emission lines studied in this paper remain at the same value as long as the galaxy has a constant average star formation rate. A galaxy with ongoing star formation continuously form new massive stars -- responsible for exciting the Balmer emission lines -- which then keeps the average H$\alpha$ and H$\beta$ luminosities stable. On the other hand, as the galaxy ages, the stars that formed earlier, with intermediate masses, remain alive. This implies an average continuum flux density that is steadily increasing with time and consequently -- since the average emission line luminosity remains constant -- lowers the mode of the EW distributions over time.

As the stochastic effects described in fig.~\ref{SED_effects} are almost entirely due to ICMF-sampling, and not IMF-sampling, their impact on the observables of dwarf galaxies at the high-redshift frontier critically hinge on the assumption that star formation primarily takes place in finite units (star clusters), and on the the assumed properties of the star clusters forming (slope and extent of the ICMF). A higher value for the slope (i.e., $\beta >-2$) introduces more high-mass clusters and therefore boosts ICMF-sampling effects and dampens IMF-sampling, and vice versa for a lower $\beta$ which suppresses the formation of high-mass clusters. The same effects are achieved by allowing a wider cluster mass-range. Assuming that not all stars form in clusters (i.e., $f_c \neq 1$, where $f_c$ is the fraction of stars formed in clusters) will decrease the impact of stochasticity as it will reduce the fluctuations in the SFR while all stars formed outside clusters will acquire masses according to the analytical IMF, given that the total mass of the galaxy is large enough \citep[i.e., $(1-f_c) \times M_\mathrm{{tot,stars}} \gtrsim 10^4 \ M_\odot$;][]{daSilva12}. For instance, any fraction of diffuse, unclustered star formation will render the relation between integrated stellar population properties and the emerging photometric and spectroscopic properties of these low-SFR galaxies tighter. This allows for the possibility, at least in principle, to use the observed dispersion in the spectral properties of high-redshift dwarf galaxies to constrain the star cluster formation in the early Universe, although this requires that other sources of scatter (most notably the burstiness of star formation) are under control. Interestingly, there are already indications from the low-redshift Universe that the observed distributions of the $L_{\mathrm{H}\alpha}/L_\mathrm{UV}$ among dwarf galaxies remains difficult to explain using standard assumptions on the ICMF and IMF when modelling star formation sampling effects \citep{Emami18}.

Throughout this paper, we have modelled the SFR fluctuations that result from star formation sampling effects in the case where the mean SFR($t$) remains constant for 100 Myr. However, low-mass galaxies at $z\gtrsim 6$ are expected to display significant SFR variations on timescales ranging from $\sim 10^6$--$10^8$ yr due to a combination of several effects (\citealt{Hopkins14,Hopkins2018}) -- the collapse of molecular clouds and star cluster formation on the shortest scales, radiative/supernova feedback on intermediate timescales and mergers and global gravitational instabilities on the longest timescales. Hence, our results simply indicate the {\it minimum} level of fluctuations one can expect at the detection limit of \textit{JWST}. In a realistic treatment, SFR fluctuations due to star formation sampling effects should be modelled simultaneously with these other sources of star formation activity variations. While some of the highest-resolution simulations already resolve mass scales in the star cluster regime \citep[e.g.,][]{Ma15}, attempts to add star formation sampling effects using \textsc{slug} at the subresolution level have been made and has been shown to contribute significantly to the variation of observable quantities \citep{Sparre17}\footnote{Note that the variations simulated with \textsc{slug} in \cite{Sparre17} come from ICMF- and IMF-sampling combined. Based on the findings in this paper, these variations originates mainly from ICMF-sampling, not IMF-sampling.}.

Furthermore we have shown that $Q_\mathrm{HeII}/Q_\mathrm{HI}$, in contrast to the other observables investigated in this paper, is affected more by IMF-sampling. The effects of stochastic star formation on $Q_\mathrm{HeII}/Q_\mathrm{HI}$ can mimic what one expects to see from low-metallicity galaxies containing a fraction of Pop III stars. In most of the cases, the extent of the variations that are due to stochasticity are deemed not significant enough to risk directly interpreting a very low-metallicity Pop II galaxy (with no detectable metal lines) suffering from sampling issues as a pure Pop III galaxy, which would require a $Q_\mathrm{HeII}/Q_\mathrm{HI}$ ratio considerably higher. Nevertheless, IMF and ICMF-sampling may still be very relevant when interpreting HeII emission lines in future \textit{JWST} observations of very faint, lensed, low-metallicity galaxies containing a smaller fraction of Pop III stars.

\section*{Acknowledgements}
AV and EZ acknowledge funding from the Swedish National Space Board.

The authors would like to thank Mark Krumholz for a helpful correspondence on the \textsc{slug} code.




\bibliographystyle{mnras}
\bibliography{bibliography} 

\begin{thebibliography}{}
\makeatletter
\relax
\def\mn@urlcharsother{\let\do\@makeother \do\$\do\&\do\#\do\^\do\_\do\%\do\~}
\def\mn@doi{\begingroup\mn@urlcharsother \@ifnextchar [ {\mn@doi@}
  {\mn@doi@[]}}
\def\mn@doi@[#1]#2{\def\@tempa{#1}\ifx\@tempa\@empty \href
  {http://dx.doi.org/#2} {doi:#2}\else \href {http://dx.doi.org/#2} {#1}\fi
  \endgroup}
\def\mn@eprint#1#2{\mn@eprint@#1:#2::\@nil}
\def\mn@eprint@arXiv#1{\href {http://arxiv.org/abs/#1} {{\tt arXiv:#1}}}
\def\mn@eprint@dblp#1{\href {http://dblp.uni-trier.de/rec/bibtex/#1.xml}
  {dblp:#1}}
\def\mn@eprint@#1:#2:#3:#4\@nil{\def\@tempa {#1}\def\@tempb {#2}\def\@tempc
  {#3}\ifx \@tempc \@empty \let \@tempc \@tempb \let \@tempb \@tempa \fi \ifx
  \@tempb \@empty \def\@tempb {arXiv}\fi \@ifundefined
  {mn@eprint@\@tempb}{\@tempb:\@tempc}{\expandafter \expandafter \csname
  mn@eprint@\@tempb\endcsname \expandafter{\@tempc}}}

\bibitem[\protect\citeauthoryear{{Applebaum}, {Brooks}, {Quinn}  \&
  {Christensen}}{{Applebaum} et~al.}{2020}]{Applebaum18}
{Applebaum} E.,  {Brooks} A.~M.,  {Quinn} T.~R.,   {Christensen} C.~R.,  2020,
  \mn@doi [\mnras] {10.1093/mnras/stz3331}, \href
  {https://ui.adsabs.harvard.edu/abs/2020MNRAS.492....8A} {492, 8}

\bibitem[\protect\citeauthoryear{{Berg}, {Erb}, {Auger}, {Pettini}  \&
  {Brammer}}{{Berg} et~al.}{2018}]{Berg2018}
{Berg} D.~A.,  {Erb} D.~K.,  {Auger} M.~W.,  {Pettini} M.,   {Brammer} G.~B.,
  2018, \mn@doi [\apj] {10.3847/1538-4357/aab7fa}, \href
  {https://ui.adsabs.harvard.edu/abs/2018ApJ...859..164B} {859, 164}

\bibitem[\protect\citeauthoryear{{Boquien}, {Buat}  \& {Perret}}{{Boquien}
  et~al.}{2014}]{Boquien14}
{Boquien} M.,  {Buat} V.,   {Perret} V.,  2014, \mn@doi [\aap]
  {10.1051/0004-6361/201424441}, \href
  {http://adsabs.harvard.edu/abs/2014A%26A...571A..72B} {571, A72}

\bibitem[\protect\citeauthoryear{{Cervi{\~n}o} \& {Luridiana}}{{Cervi{\~n}o} \&
  {Luridiana}}{2004}]{Cervino04}
{Cervi{\~n}o} M.,  {Luridiana} V.,  2004, \mn@doi [\aap]
  {10.1051/0004-6361:20031454}, \href
  {http://adsabs.harvard.edu/abs/2004A%26A...413..145C} {413, 145}

\bibitem[\protect\citeauthoryear{{Eldridge}}{{Eldridge}}{2012}]{Eldridge12}
{Eldridge} J.~J.,  2012, \mn@doi [\mnras] {10.1111/j.1365-2966.2012.20662.x},
  \href {http://adsabs.harvard.edu/abs/2012MNRAS.422..794E} {422, 794}

\bibitem[\protect\citeauthoryear{{Elmegreen}}{{Elmegreen}}{2000}]{Elmegreen2000}
{Elmegreen} B.~G.,  2000, \mn@doi [\apj] {10.1086/309204}, \href
  {https://ui.adsabs.harvard.edu/\#abs/2000ApJ...539..342E} {539, 342}

\bibitem[\protect\citeauthoryear{{Emami}, {Siana}, {Weisz}, {Johnson}, {Ma}  \&
  {El-Badry}}{{Emami} et~al.}{2019}]{Emami18}
{Emami} N.,  {Siana} B.,  {Weisz} D.~R.,  {Johnson} B.~D.,  {Ma} X.,
  {El-Badry} K.,  2019, \mn@doi [\apj] {10.3847/1538-4357/ab211a}, \href
  {https://ui.adsabs.harvard.edu/abs/2019ApJ...881...71E} {881, 71}

\bibitem[\protect\citeauthoryear{{Fall}, {Krumholz}  \& {Matzner}}{{Fall}
  et~al.}{2010}]{Fall2010}
{Fall} S.~M.,  {Krumholz} M.~R.,   {Matzner} C.~D.,  2010, \mn@doi [\apj]
  {10.1088/2041-8205/710/2/L142}, \href
  {https://ui.adsabs.harvard.edu/abs/2010ApJ...710L.142F} {710, L142}

\bibitem[\protect\citeauthoryear{{Forero-Romero} \& {Dijkstra}}{{Forero-Romero}
  \& {Dijkstra}}{2013}]{ForeroRomero13}
{Forero-Romero} J.~E.,  {Dijkstra} M.,  2013, \mn@doi [\mnras]
  {10.1093/mnras/sts177}, \href
  {http://adsabs.harvard.edu/abs/2013MNRAS.428.2163F} {428, 2163}

\bibitem[\protect\citeauthoryear{{Fouesneau} \& {Lan{\c c}on}}{{Fouesneau} \&
  {Lan{\c c}on}}{2010}]{Fousneau10}
{Fouesneau} M.,  {Lan{\c c}on} A.,  2010, \mn@doi [\aap]
  {10.1051/0004-6361/201014084}, \href
  {http://adsabs.harvard.edu/abs/2010A%26A...521A..22F} {521, A22}

\bibitem[\protect\citeauthoryear{{Fumagalli}, {da Silva}  \&
  {Krumholz}}{{Fumagalli} et~al.}{2011}]{Fumagalli11}
{Fumagalli} M.,  {da Silva} R.~L.,   {Krumholz} M.~R.,  2011, \mn@doi [\apjl]
  {10.1088/2041-8205/741/2/L26}, \href
  {http://adsabs.harvard.edu/abs/2011ApJ...741L..26F} {741, L26}

\bibitem[\protect\citeauthoryear{{Hernandez}}{{Hernandez}}{2012}]{Hernandez12}
{Hernandez} X.,  2012, \mn@doi [\mnras] {10.1111/j.1365-2966.2011.20099.x},
  \href {http://adsabs.harvard.edu/abs/2012MNRAS.420.1183H} {420, 1183}

\bibitem[\protect\citeauthoryear{{Hopkins}, {Kere{\v{s}}}, {O{\~n}orbe},
  {Faucher-Gigu{\`e}re}, {Quataert}, {Murray}  \& {Bullock}}{{Hopkins}
  et~al.}{2014}]{Hopkins14}
{Hopkins} P.~F.,  {Kere{\v{s}}} D.,  {O{\~n}orbe} J.,  {Faucher-Gigu{\`e}re}
  C.-A.,  {Quataert} E.,  {Murray} N.,   {Bullock} J.~S.,  2014, \mn@doi
  [\mnras] {10.1093/mnras/stu1738}, \href
  {https://ui.adsabs.harvard.edu/abs/2014MNRAS.445..581H} {445, 581}

\bibitem[\protect\citeauthoryear{{Hopkins} et~al.,}{{Hopkins}
  et~al.}{2018}]{Hopkins2018}
{Hopkins} P.~F.,  et~al., 2018, \mn@doi [\mnras] {10.1093/mnras/sty1690}, \href
  {https://ui.adsabs.harvard.edu/abs/2018MNRAS.480..800H} {480, 800}

\bibitem[\protect\citeauthoryear{{Inoue}}{{Inoue}}{2011}]{Inoue11}
{Inoue} A.~K.,  2011, \mn@doi [\mnras] {10.1111/j.1365-2966.2011.18906.x},
  \href {https://ui.adsabs.harvard.edu/abs/2011MNRAS.415.2920I} {415, 2920}

\bibitem[\protect\citeauthoryear{{Kikuchihara} et~al.,}{{Kikuchihara}
  et~al.}{2019}]{Kikuchihara19}
{Kikuchihara} S.,  et~al., 2019, arXiv e-prints, \href
  {https://ui.adsabs.harvard.edu/abs/2019arXiv190506927K} {p. arXiv:1905.06927}

\bibitem[\protect\citeauthoryear{{Kroupa}}{{Kroupa}}{2001}]{kroupa2001}
{Kroupa} P.,  2001, \mn@doi [\mnras] {10.1046/j.1365-8711.2001.04022.x}, \href
  {https://ui.adsabs.harvard.edu/abs/2001MNRAS.322..231K} {322, 231}

\bibitem[\protect\citeauthoryear{{Krumholz}, {Fumagalli}, {da Silva}, {Rendahl}
   \& {Parra}}{{Krumholz} et~al.}{2015}]{Krumholz15}
{Krumholz} M.~R.,  {Fumagalli} M.,  {da Silva} R.~L.,  {Rendahl} T.,   {Parra}
  J.,  2015, \mn@doi [\mnras] {10.1093/mnras/stv1374}, \href
  {https://ui.adsabs.harvard.edu/abs/2015MNRAS.452.1447K} {452, 1447}

\bibitem[\protect\citeauthoryear{{Krumholz}, {Adamo}, {Fumagalli}  \&
  {Calzetti}}{{Krumholz} et~al.}{2019}]{Krumholz19}
{Krumholz} M.~R.,  {Adamo} A.,  {Fumagalli} M.,   {Calzetti} D.,  2019, \mn@doi
  [\mnras] {10.1093/mnras/sty2896}, \href
  {http://adsabs.harvard.edu/abs/2019MNRAS.482.3550K} {482, 3550}

\bibitem[\protect\citeauthoryear{{Lada} \& {Lada}}{{Lada} \&
  {Lada}}{2003}]{LadaLada2003}
{Lada} C.~J.,  {Lada} E.~A.,  2003, \mn@doi [\araa]
  {10.1146/annurev.astro.41.011802.094844}, \href
  {https://ui.adsabs.harvard.edu/abs/2003ARA&A..41...57L} {41, 57}

\bibitem[\protect\citeauthoryear{{Ma}, {Kasen}, {Hopkins},
  {Faucher-Gigu{\`e}re}, {Quataert}, {Kere{\v s}}  \& {Murray}}{{Ma}
  et~al.}{2015}]{Ma15}
{Ma} X.,  {Kasen} D.,  {Hopkins} P.~F.,  {Faucher-Gigu{\`e}re} C.-A.,
  {Quataert} E.,  {Kere{\v s}} D.,   {Murray} N.,  2015, \mn@doi [\mnras]
  {10.1093/mnras/stv1679}, \href
  {http://adsabs.harvard.edu/abs/2015MNRAS.453..960M} {453, 960}

\bibitem[\protect\citeauthoryear{{Madau} \& {Dickinson}}{{Madau} \&
  {Dickinson}}{2014}]{Madau14}
{Madau} P.,  {Dickinson} M.,  2014, \mn@doi [\araa]
  {10.1146/annurev-astro-081811-125615}, \href
  {https://ui.adsabs.harvard.edu/abs/2014ARA&A..52..415M} {52, 415}

\bibitem[\protect\citeauthoryear{{Mas-Ribas}, {Dijkstra}  \&
  {Forero-Romero}}{{Mas-Ribas} et~al.}{2016}]{MasRibas16}
{Mas-Ribas} L.,  {Dijkstra} M.,   {Forero-Romero} J.~E.,  2016, \mn@doi [\apj]
  {10.3847/1538-4357/833/1/65}, \href
  {http://adsabs.harvard.edu/abs/2016ApJ...833...65M} {833, 65}

\bibitem[\protect\citeauthoryear{{{\"O}stlin}, {Zackrisson}, {Sollerman},
  {Mattila}  \& {Hayes}}{{{\"O}stlin} et~al.}{2008}]{ostlin08}
{{\"O}stlin} G.,  {Zackrisson} E.,  {Sollerman} J.,  {Mattila} S.,   {Hayes}
  M.,  2008, \mn@doi [\mnras] {10.1111/j.1365-2966.2008.13319.x}, \href
  {https://ui.adsabs.harvard.edu/abs/2008MNRAS.387.1227O} {387, 1227}

\bibitem[\protect\citeauthoryear{{Paalvast} \& {Brinchmann}}{{Paalvast} \&
  {Brinchmann}}{2017}]{Paalvast17}
{Paalvast} M.,  {Brinchmann} J.,  2017, \mn@doi [\mnras]
  {10.1093/mnras/stx1271}, \href
  {https://ui.adsabs.harvard.edu/\#abs/2017MNRAS.470.1612P} {470, 1612}

\bibitem[\protect\citeauthoryear{{Planck Collaboration} et~al.,}{{Planck
  Collaboration} et~al.}{2016}]{Planck16}
{Planck Collaboration} et~al., 2016, \mn@doi [\aap]
  {10.1051/0004-6361/201525830}, \href
  {https://ui.adsabs.harvard.edu/abs/2016A&A...594A..13P} {594, A13}

\bibitem[\protect\citeauthoryear{{Raiter}, {Schaerer}  \& {Fosbury}}{{Raiter}
  et~al.}{2010}]{raiter10}
{Raiter} A.,  {Schaerer} D.,   {Fosbury} R.~A.~E.,  2010, \mn@doi [\aap]
  {10.1051/0004-6361/201015236}, \href
  {https://ui.adsabs.harvard.edu/abs/2010A&A...523A..64R} {523, A64}

\bibitem[\protect\citeauthoryear{{Santos} \& {Frogel}}{{Santos} \&
  {Frogel}}{1997}]{Santos97}
{Santos} Jr. J.~F.~C.,  {Frogel} J.~A.,  1997, \mn@doi [\apj] {10.1086/303921},
  \href {http://adsabs.harvard.edu/abs/1997ApJ...479..764S} {479, 764}

\bibitem[\protect\citeauthoryear{{Sarmento}, {Scannapieco}  \&
  {Cohen}}{{Sarmento} et~al.}{2018}]{Sarmento2018}
{Sarmento} R.,  {Scannapieco} E.,   {Cohen} S.,  2018, \mn@doi [\apj]
  {10.3847/1538-4357/aa989a}, \href
  {https://ui.adsabs.harvard.edu/abs/2018ApJ...854...75S} {854, 75}

\bibitem[\protect\citeauthoryear{{Sarmento}, {Scannapieco}  \&
  {C{\^o}t{\'e}}}{{Sarmento} et~al.}{2019}]{Sarmento2019}
{Sarmento} R.,  {Scannapieco} E.,   {C{\^o}t{\'e}} B.,  2019, \mn@doi [\apj]
  {10.3847/1538-4357/aafa1a}, \href
  {https://ui.adsabs.harvard.edu/abs/2019ApJ...871..206S} {871, 206}

\bibitem[\protect\citeauthoryear{{Schaerer}}{{Schaerer}}{2003}]{Schaerer2003}
{Schaerer} D.,  2003, \mn@doi [\aap] {10.1051/0004-6361:20021525}, \href
  {https://ui.adsabs.harvard.edu/abs/2003A&A...397..527S} {397, 527}

\bibitem[\protect\citeauthoryear{{Schaerer}, {Fragos}  \& {Izotov}}{{Schaerer}
  et~al.}{2019}]{Schaerer2019}
{Schaerer} D.,  {Fragos} T.,   {Izotov} Y.~I.,  2019, \mn@doi [\aap]
  {10.1051/0004-6361/201935005}, \href
  {https://ui.adsabs.harvard.edu/abs/2019A&A...622L..10S} {622, L10}

\bibitem[\protect\citeauthoryear{{Shimizu}, {Inoue}, {Okamoto}  \&
  {Yoshida}}{{Shimizu} et~al.}{2014}]{Shimizu14}
{Shimizu} I.,  {Inoue} A.~K.,  {Okamoto} T.,   {Yoshida} N.,  2014, \mn@doi
  [\mnras] {10.1093/mnras/stu265}, \href
  {https://ui.adsabs.harvard.edu/abs/2014MNRAS.440..731S} {440, 731}

\bibitem[\protect\citeauthoryear{{Shirazi} \& {Brinchmann}}{{Shirazi} \&
  {Brinchmann}}{2012}]{shirazi2012}
{Shirazi} M.,  {Brinchmann} J.,  2012, \mn@doi [\mnras]
  {10.1111/j.1365-2966.2012.20439.x}, \href
  {https://ui.adsabs.harvard.edu/abs/2012MNRAS.421.1043S} {421, 1043}

\bibitem[\protect\citeauthoryear{{Shivaei} et~al.,}{{Shivaei}
  et~al.}{2015}]{Shivaei15}
{Shivaei} I.,  et~al., 2015, \mn@doi [\apj] {10.1088/0004-637X/815/2/98}, \href
  {https://ui.adsabs.harvard.edu/abs/2015ApJ...815...98S} {815, 98}

\bibitem[\protect\citeauthoryear{{Sparre}, {Hayward}, {Feldmann},
  {Faucher-Gigu{\`e}re}, {Muratov}, {Kere{\v{s}}}  \& {Hopkins}}{{Sparre}
  et~al.}{2017}]{Sparre17}
{Sparre} M.,  {Hayward} C.~C.,  {Feldmann} R.,  {Faucher-Gigu{\`e}re} C.-A.,
  {Muratov} A.~L.,  {Kere{\v{s}}} D.,   {Hopkins} P.~F.,  2017, \mn@doi
  [\mnras] {10.1093/mnras/stw3011}, \href
  {https://ui.adsabs.harvard.edu/abs/2017MNRAS.466...88S} {466, 88}

\bibitem[\protect\citeauthoryear{{Yue} et~al.,}{{Yue} et~al.}{2018}]{Yue18}
{Yue} B.,  et~al., 2018, \mn@doi [\apj] {10.3847/1538-4357/aae77f}, \href
  {https://ui.adsabs.harvard.edu/abs/2018ApJ...868..115Y} {868, 115}

\bibitem[\protect\citeauthoryear{{Zackrisson}, {Bergvall}  \&
  {{\"O}stlin}}{{Zackrisson} et~al.}{2005}]{Zackrisson05}
{Zackrisson} E.,  {Bergvall} N.,   {{\"O}stlin} G.,  2005, \mn@doi [\aap]
  {10.1051/0004-6361:20041585}, \href
  {https://ui.adsabs.harvard.edu/abs/2005A&A...435...29Z} {435, 29}

\bibitem[\protect\citeauthoryear{{Zackrisson}, {Inoue}  \&
  {Jensen}}{{Zackrisson} et~al.}{2013}]{zackrisson13}
{Zackrisson} E.,  {Inoue} A.~K.,   {Jensen} H.,  2013, \mn@doi [\apj]
  {10.1088/0004-637X/777/1/39}, \href
  {https://ui.adsabs.harvard.edu/abs/2013ApJ...777...39Z} {777, 39}

\bibitem[\protect\citeauthoryear{{Zackrisson} et~al.,}{{Zackrisson}
  et~al.}{2017}]{Zackrisson2017}
{Zackrisson} E.,  et~al., 2017, \mn@doi [\apj] {10.3847/1538-4357/836/1/78},
  \href {https://ui.adsabs.harvard.edu/abs/2017ApJ...836...78Z} {836, 78}

\bibitem[\protect\citeauthoryear{{da Silva}, {Fumagalli}  \& {Krumholz}}{{da
  Silva} et~al.}{2011}]{daSilva11}
{da Silva} R.~L.,  {Fumagalli} M.,   {Krumholz} M.,  2011, {SLUG:
  Stochastically Lighting Up Galaxies} (\mn@eprint {ascl} {1106.012})

\bibitem[\protect\citeauthoryear{{da Silva}, {Fumagalli}  \& {Krumholz}}{{da
  Silva} et~al.}{2012}]{daSilva12}
{da Silva} R.~L.,  {Fumagalli} M.,   {Krumholz} M.,  2012, \mn@doi [\apj]
  {10.1088/0004-637X/745/2/145}, \href
  {http://adsabs.harvard.edu/abs/2012ApJ...745..145D} {745, 145}

\makeatother
\end{thebibliography}







\bsp	
\label{lastpage}
\end{document}